\documentclass[preprint,grl]{agu2001}

\usepackage{amsfonts}
\usepackage{amsmath}
\usepackage{graphicx}

\newcommand{\coloneqq}{\ensuremath{\mathrel{:=}}}

\newcommand{\Kb}{\emph{K. brevis }}

\authorrunninghead{OLASCOAGA ET AL.}
\titlerunninghead{Persistent Transport Barrier on the West Florida Shelf}
\lefthead{OLASCOAGA ET AL.} \righthead{Persistent Transport Barrier
on the West Florida Shelf} \journalid{2006} \articleid{X}{X}
\paperid{2006XXXXXXXXX} \cpright{AGU}{2006} 
\received{\today} \revised{\today} \accepted{\today}

\authoraddr{M.~J. Olascoaga, I.~I. Rypina, M.~G. Brown and F.~J. Beron-Vera,
RSMAS/AMP, University of Miami, 4600 Rickenbacker Cswy., Miami, FL
33149, USA. (jolascoaga@rsmas.miami.edu)}
\authoraddr{H. Ko\c{c}ak, Departments of Computer Science and Mathematics,
University of Miami, 1365 Memorial Dr., Coral Gables, FL 33124, USA.}
\authoraddr{L.~E. Brand, RSMAS/MBF, University of Miami, 4600 Rickenbacker Cswy.,
Miami, FL 33149, USA.}
\authoraddr{G.~R. Halliwell and L.~K. Shay, RSMAS/MPO, University of Miami,
4600 Rickenbacker Cswy., Miami, FL
33149, USA.}

\begin{document}

\title{Persistent Transport Barrier on the West Florida Shelf}

\authors{M.~J. Olascoaga, \altaffilmark{1} I.~I. Rypina, \altaffilmark{1}
M.~G. Brown, \altaffilmark{1} F.~J. Beron-Vera, \altaffilmark{1} H.
Ko\c{c}ak, \altaffilmark{2} L.~E. Brand, \altaffilmark{1} G.~R.
Halliwell \altaffilmark{1} and L.~K. Shay \altaffilmark{1}}

\altaffiltext{1}{Rosenstiel School of Marine and Atmospheric
Science, University of Miami, Miami, Florida, USA.}
\altaffiltext{2}{Departments of Computer Science and Mathematics,
University of Miami, Miami, Florida, USA.}

\begin{abstract}\baselineskip.4cm

Analysis of drifter trajectories in the Gulf of Mexico has revealed
the existence of a region on the southern portion of the West
Florida Shelf (WFS) that is not visited by drifters that are
released outside of the region. This so-called ``forbidden zone''
(FZ) suggests the existence of a persistent cross-shelf transport
barrier on the southern portion of the WFS. In this letter a
year-long record of surface currents produced by a Hybrid-Coordinate
Ocean Model simulation of the WFS is used to identify Lagrangian
coherent structures (LCSs), which reveal the presence of a robust
and persistent cross-shelf transport barrier in approximately the
same location as the boundary of the FZ. The location of the
cross-shelf transport barrier undergoes a seasonal oscillation,
being closer to the coast in the summer than in the winter. A
month-long record of surface currents inferred from high-frequency
(HF) radar measurements in a roughly 60 km $\times$ 80 km region on
the WFS off Tampa Bay is also used to identify LCSs, which reveal
the presence of robust transient transport barriers. While the
HF-radar-derived transport barriers cannot be unambiguously linked
to the boundary of the FZ, this analysis does demonstrate the
feasibility of monitoring transport barriers on the WFS using a
HF-radar-based measurement system. The implications of a persistent
cross-shelf transport barrier on the WFS for the development of
harmful algal blooms on the shoreward side of the barrier are
considered.

\end{abstract}

\begin{article}\baselineskip.4cm

\section{Introduction}

\citet{Yang-etal-99} presents the results of the analysis of
trajectories of satellite-tracked drifters released during the
period February 1996 through February 1997 on the continental shelf
in the northeastern portion of the Gulf of Mexico (GoM). Inspection
of the drifter trajectories depicted in Figure 2 of that paper
reveals the presence of a trajectory-free triangular-shaped region
on the southern portion of the West Florida Shelf (WFS), which has
been referred to by those authors as a ``forbidden zone'' (FZ).
Although little can be inferred about the spatio-temporal
variability of the FZ from the aforementioned figure, the FZ almost
certainly wobbles with a complicated spatio-temporal structure. This
expectation finds some support in the seasonal analysis of drifter
trajectories presented by \citet{Morey-etal-03}, which  also
included trajectories of drifters released on the continental shelf
in the northwestern portion of the GoM.

The presence of the FZ suggests the existence of a seemingly robust
barrier on the WFS that inhibits the transport across the shelf.
This cross-shelf transport barrier not only can have implications
for pollutant dispersal, but may also be consequential for harmful
algal blooms on the shoreward side of the barrier.

In this letter we employ methods from dynamical systems theory to
attain a twofold goal. First, we seek to demonstrate the robustness
of the suggested cross-shelf transport barrier on the WFS in the
inspection of drifter trajectories through the analysis of simulated
surface currents. Second, we seek to demonstrate the feasibility of
monitoring transport barriers on the WFS using high-frequency (HF)
radar measurements.

\begin{figure*}[t]
\centering
\includegraphics[width=39pc]{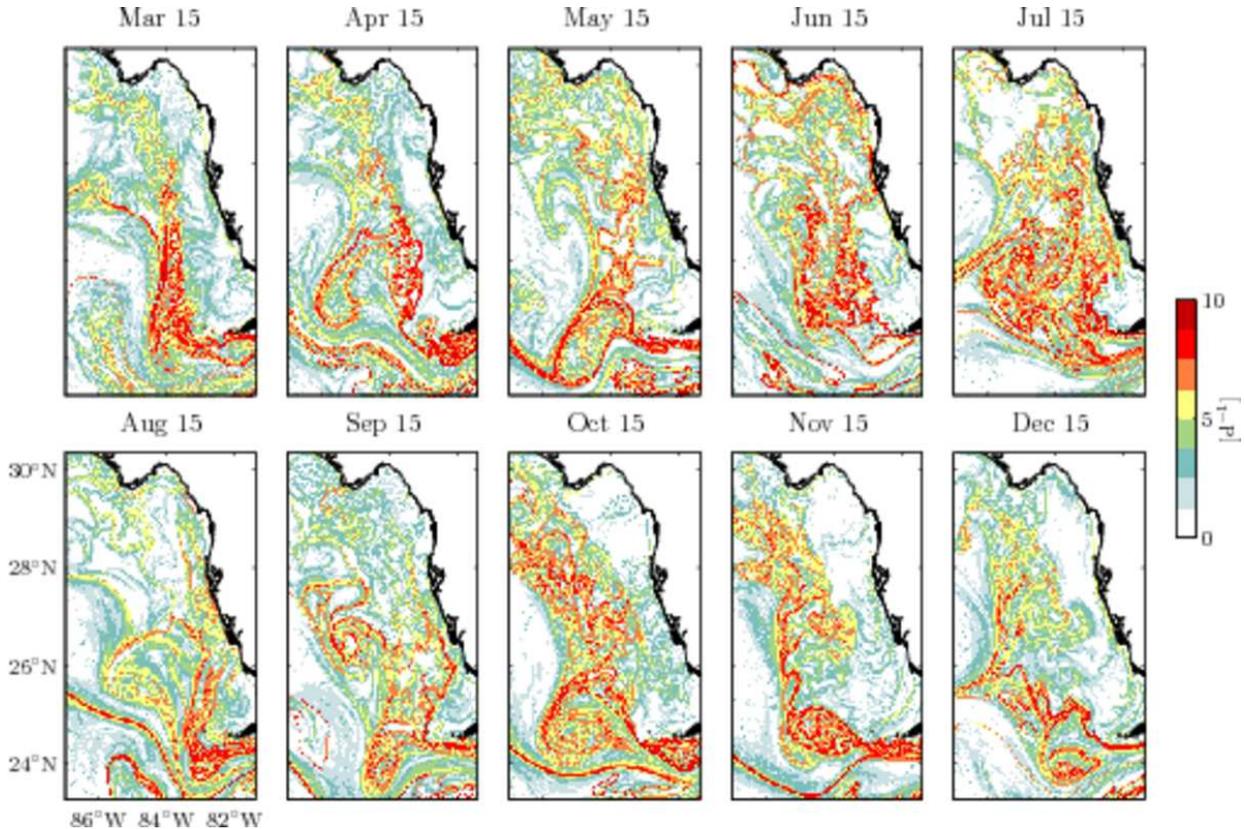}
\caption{Sequence of snapshots of FTLE field (\ref{FTLE}) computed
backwards in time using surface currents generated by a HYCOM
simulation of the WFS for year 2004. Maximizing ridges of FTLE field
define attracting LCSs, which approximate unstable manifolds and
thus constitute transport barriers. Note the presence of a
triangular-shaped area on the southern portion of the WFS with small
FTLEs bounded by the western Florida coast on the east, the lower
Florida keys on the south, and large maximizing ridges of FTLE field
on the west. The latter constitute a barrier for cross-shelf
transport, which is seen to undergo a seasonal oscillation
approximately about the western boundary of the FZ.}
\label{FTLE_HYCOM}
\end{figure*}

\section{Lagrangian Coherent Structures (LCSs)}\label{LCSs}

Theoretical work on dynamical systems
\citep[e.g.,][]{Haller-00,Haller-Yuan-00,Haller-01a,
Haller-01b,Haller-02,Shadden-etal-05} has characterized transport
barriers in unsteady two-dimensional incompressible flows as
Lagrangian coherent structures (LCSs). The theory behind LCSs will
not be discussed in this letter. We note, however, that the LCSs of
interest correspond to the stable and/or unstable invariant
manifolds of hyperbolic points.

An invariant manifold can be understood as a material curve of
fluid, i.e., composed always of the same fluid particles. Associated
with a hyperbolic point in a steady flow there are two invariant
manifolds, one stable and another one unstable. Along the stable
(unstable) manifold, a fluid particle asymptotically approaches the
hyperbolic point in forward (backward) time. Initially nearby fluid
particle trajectories flanking a stable (unstable) manifold repel
(attract) from each other at an exponential rate. These manifolds
therefore constitute separatrices that divide the space into regions
with dynamically distinct characteristics. Furthermore, being
material curves these separatrices cannot be traversed by fluid
particles, i.e., they constitute true transport barriers. In an
unsteady flow there are also hyperbolic points with associated
stable and unstable manifolds. Unlike the steady case, these
hyperbolic points are not still but rather undergo motion, which is
typically aperiodic in predominantly turbulent ocean flows. As in
the steady case, the associated manifolds also constitute
separatrices, and hence transport barriers, albeit in a spatially
local sense and for sufficiently short time. The latter is seen in
that generically there is chaotic motion in the vicinity of the
points where the stable and unstable manifolds intersect one another
after successively stretching and folding. These intersections lead
to the formation of regions with a highly intricate tangle-like
structure. In these regions trajectories of initially nearby fluid
particles rapidly diverge and fluid particles from other regions are
injected in between, which constitutes a very effective mechanism
for mixing.

Identification of LCSs, which is impossible from naked-eye
inspection of snapshots of simulated or measured velocity fields and
at most very difficult from the inspection of individual fluid
particle trajectories, is critically important for understanding
transport and mixing of tracers in the ocean. The computation of
finite-time Lyapunov exponents (FTLEs) provides practical means for
identifying repelling and attracting LCSs that approximate stable
and unstable manifolds, respectively. The FTLE is the finite-time
average of the maximum expansion or contraction rate for pairs of
passively advected fluid particles. More specifically, the FTLE is
defined as
\begin{equation}
   \sigma_{t}^{\tau}(\mathbf{x}) \coloneqq \frac{1}{|\tau|}
   \ln\|\partial_{\mathbf{x}}
   \varphi_{t}^{t+\tau}(\mathbf{x})\|, \label{FTLE}
\end{equation}
where $\left\Vert \, \right\Vert $ denotes the L$_{2}$ norm and
$\varphi _{t}^{t+\tau}:\mathbf{x}(t)\mapsto \mathbf{x}(t+\tau)$ is
the flow map that takes fluid particles from their initial location
at time $t$ to their location at time $t+\tau.$ The flow map
$\varphi _{t}^{t+\tau}$ is obtained by solving the fluid particle
motion regarded as a dynamical system obeying
\begin{equation}
   \dot{\mathbf{x}} = \mathbf{u}(\mathbf{x},t), \label{dxdt}
\end{equation}
where the overdot stands for time differentiation and
$\mathbf{u}(\mathbf{x},t)$ is a prescribed velocity field. Repelling
and attracting LCSs are then defined
\citep{Haller-02,Shadden-etal-05} as maximizing ridges of FTLE field
computed forward ($\tau > 0$) and backward ($\tau < 0$) in time,
respectively.

We remark that while LCSs are fundamentally a Lagrangian concept,
the algorithm used to identify such structures requires a high
resolution Eulerian description of the flow. Recently,
\citet{Lekien-etal-05} has successfully applied this theory to
identify transport barriers using HF-radar-derived surface velocity
in the east Florida coast.

\section{LCSs Derived from Simulated Currents}\label{HYCOM}

Numerical model output provides a flow description
$\mathbf{u}(\mathbf{x},t)$ that is suitable for use to identify
LCSs. Also, it has the advantage of allowing for a spatio-temporal
coverage that is impossible to attain with existing observational
systems. Here we consider a year-long record of surface currents
produced by a Hybrid-Coordinate Ocean Model (HYCOM) simulation along
the WFS for year 2004.

The year-long record of simulated currents consists of daily surface
velocity fields extracted in the WFS domain from a
0.04$^{\circ}$-resolution, free-running HYCOM simulation of the GoM,
itself nested within a 0.08$^{\circ}$-resolution Atlantic basin data
assimilative nowcast, which was generated at the Naval Research
Laboratory as part of a National Oceanographic Partnership Program
in support of the Global Ocean Data Assimilation Experiment
\citep{Chassignet-etal-06a,Chassignet-etal-06b}. The Atlantic
nowcast was forced with realistic high-frequency forcing obtained
from the U. S. Navy NOGAPS operational atmospheric model. It
assimilated sea surface temperature and anomalous sea surface height
from satellite altimetry with downward projection of anomalous
temperature and salinity profiles. The nested GoM model was
free-running and driven by the same high-frequency atmospheric
forcing.  The topography used in both models was derived from the
ETOPO5 dataset, with the coastline in the GoM model following the 2
m isobath.  Both models included river runoff.

Figure \ref{FTLE_HYCOM} shows snapshots of FTLE field, which were
computed using the software package MANGEN, a dynamical systems
toolkit designed by F. Lekien that is available at
http://www.lekien.com/$\sim$francois/software/mangen. At each time
$t$ the algorithm coded in MANGEN performs the following tasks.
First, system (\ref{dxdt}) is integrated using a fourth-order
Runge--Kutta--Fehlberg method for a grid of particles at time $t$ to
get their positions at time $t+\tau$, which are the values of the
flow map at each point. This requires interpolating the velocity
data, which is carried out employing a cubic method. Second, the
spatial gradient of the flow map is obtained at each point in the
initial grid by central differencing with neighboring grid points.
Third, the FTLE is computed at each point in the initial grid by
evaluating (\ref{FTLE}). The previous three steps are repeated for a
range of $t$ values to produce a time series of FTLE field. Here we
have set $\tau = - 60$ d so that the maximizing ridges of the FTLE
fields shown in Figure \ref{FTLE_HYCOM} correspond to attracting
LCSs. The choice $\tau = -60$ d was suggested by the time it takes a
typical fluid particle to leave the WFS domain. Clearly, some
particles will exit the domain before 60 d of integration. In this
case, MANGEN evaluates expression (\ref{FTLE}) using the position of
each such particles prior exiting the domain. Note that due to the
choice $\tau = -60$ d the time series of computed FTLE fields based
on our year-long record of simulated currents can only have a
10-month maximum duration.

The regions of most intense red tonalities in each panel of Figure
\ref{FTLE_HYCOM} roughly indicate maximizing ridges of FTLE field.
These regions are seen to form smooth, albeit highly structured,
curves that constitute the desired LCSs or transport barriers. Of
particular interest is the triangular-shaped area on the southern
portion of the WFS with small FTLEs bounded by the western Florida
coast on the east, the lower Florida keys on the south, and large
maximizing ridges of FTLE field on the west. The latter constitute a
cross-shelf transport barrier that approximately coincides in
position with the western boundary of the FZ identified in
\citet{Yang-etal-99}. This is most clearly visible during the period
May through September. The sequence of snapshots of FTLE field in
Figure \ref{FTLE_HYCOM} also reveals a seasonal movement of the
cross-shelf transport barrier, being offshore during the winter and
closer to the coast during the summer, which is in agreement with
drifter observations \citep{Morey-etal-03}.

\begin{figure*}[t]
\centering
\includegraphics[width=31pc]{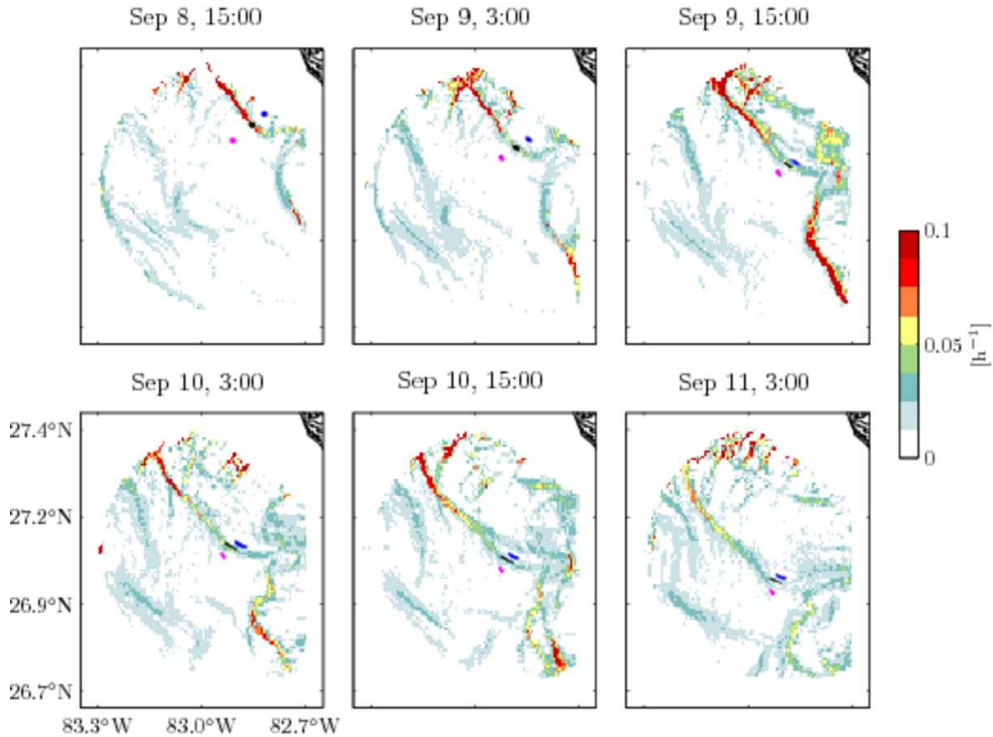}
\caption{As in Figure \ref{FTLE_HYCOM} but using HF-radar-derived
surface currents in a roughly 60 km $\times$ 80 km domain on the WFS
off Tampa Bay. The black, dark-blue, and magenta spots in each panel
indicate clusters of passively advected fluid particles tracked from
their initial location in the upper-left panel. Note how the black
spot stretches along one of the identified LCSs and the dark-blue
and magenta spots flanking this LCS are attracted to the LCS but do
not cross it.} \label{FTLE_HF}
\end{figure*}

\section{LCSs Derived from Measured Currents}\label{HF}

Figure \ref{FTLE_HF} shows a sequence of snapshots of FTLE field
computed using surface velocity inferred from HF radar measurements
taken during September 2003 in an approximately 60 km $\times$ 80 km
domain on the WFS off Tampa Bay.

The HF radar measurements consist of measurements made with Wellen
radars, which mapped coastal ocean surface currents over the above
domain with 30-minute intervals \citep{Shay-etal-06}. The
computation of FTLEs was performed backward in time ($\tau = -60$ h)
so that the maximizing ridges of FTLE field in Figure \ref{FTLE_HF}
indicate attracting LCSs, which are analogous to perturbed unstable
invariant manifolds. The numerical computation of the FTLEs was not
carried out using the MANGEN software package. Instead, we developed
our own MATLAB codes, which, employing a methodology similar to that
outlined in the previous section, allowed us to more easily handling
FTLE computation based on velocity data defined on an irregular and
totally open domain.

The transport barrier character of the attracting LCSs identified in
the FTLE fields depicted in Figure \ref{FTLE_HF} is illustrated by
numerically simulating the evolution of three clusters of fluid
particles. One of the clusters (black spot in the figure) is
initially chosen on top of one LCS, while the other two clusters
(dark-blue and magenta spots in the figure) are initially located at
one side and the other of the LCS. Looking at the evolution of the
clusters in time, we note that the black cluster remains on top of
the LCS and stretches along it as time progresses. Also note that
dark-blue and magenta clusters remain on two different sides,
indicating the absence of flux across the LCS.

We remark that LCSs are identifiable in the region covered by the HF
radar system through the whole month of September 2003. However,
because of the limited domain of the radar footprint and the short
deployment time window, we cannot say with certainty that any of the
observed structures corresponds to the boundary of the FZ. In spite
of this uncertainty, our analysis of the HF radar measurements is
highly encouraging inasmuch as it demonstrates the feasibility of
tracking the evolution of the boundary of the FZ in near real time.

\section{Biological Implications}

In addition to being an interesting physical feature whose
underlying dynamics deserves further study, the cross-shelf
transport barrier on the WFS also has potentially important
biological implications.

The toxic dinoflagellate \emph{Karenia brevis} has a rather
restricted spatial distribution, primarily the GoM
\citep{Kusek-etal-99}. While \Kb exists in low concentrations in
vast areas of the GoM, it occasionally forms blooms along the
northern and eastern coasts of the GoM
\citep{Wilson-Ray-56,Geesey-Tester-93,Tester-Steidinger-97,%
Dortch-etal-98,Kusek-etal-99,Magana-etal-03}. The largest and most
frequent blooms, however, occur along the southern portion of the
WFS \citep{Steidinger-Haddad-81,Kusek-etal-99}. Where these occur,
there can be widespread death of fish, manatees, dolphins, turtles,
and seabirds as a result of the brevetoxins produced by \Kb
\citep{Bossart-etal-98,Landsberg-Steidinger-98,Landsberg-02,%
Shumway-etal-03,Flewelling-etal-05}. The brevetoxins also end up in
an aerosol, which affects human respiration
\citep{Backer-etal-03,Kirkpatrick-etal-04}. Being slow growing while
fast swimming algae \citep{Chan-78,Brand-Guillard-81},
dinoflagellates have the highest potential for achieving high
population densities as a consequence of biophysical accumulation
rather than actual growth. As a result of expatriate losses, \Kb
(and other dinoflagellates) would be expected to develop large
blooms only in regions where stirring by currents is low. Indeed,
many dinoflagellate blooms tend to occur in enclosed basins such as
estuaries and lagoons, where expatriate losses are reduced. However,
as mentioned above, \Kb often forms large blooms along the open
coastline of the southern portion of the WFS, typically inside the
FZ.

We hypothesize that the cross-shelf transport barrier associated
with the FZ provides a mechanism that reduces \Kb expatriation. A
corollary of this hypothesis is that this barrier allows the
nutrients from land runoff to accumulate near the coastline rather
than being swept away by currents. While we cannot explain why \Kb
often dominates over other species of dinoflagellates in the FZ, we
can predict that slow growing dinoflagellates will be more prevalent
within the FZ than outside.

\section{Summary}

In this letter we have shown that, when analyzed using dynamical
systems methods, a year-long record of surface currents produced by
a regional version of HYCOM reveals the presence of a cross-shelf
transport barrier on the southern portion of the WFS, which is in
approximately the same location as the boundary of the FZ identified
earlier by \citet{Yang-etal-99} using satellite-tracked drifter
trajectories. The simulated cross-shelf transport barrier was
robust, being present in all seasons while undergoing a seasonal
oscillation. The simulated cross-shelf transport barrier was closer
to shore in the summer months than in the winter months in agreement
with observations \citep{Morey-etal-03}. HF radar measurements were
analyzed in a similar fashion and this analysis demonstrated the
feasibility of experimentally monitoring transport barriers on the
WFS using a system that can be operated in near real time.

\begin{acknowledgments}
We thank I. Udovydchenkov for useful discussions. MJO was supported
by NSF grant CMG-0417425 and PARADIGM NSF/ONR-NOPP grant
N000014-02-1-0370. IAR, MGB, FJBV, and HK were supported by NSF
grant CMG-0417425. LKS and the WERA HF Radar deployment and analyses
were supported by ONR grant N00014-02-1-0972 through the SEA-COOS
program administered by UNC-CH. The NOAA ACT program provided
partial travel support for the WERA deployment personnel. We thank
T. Cook, B. Haus, J. Martinez, and S. Guhin in deploying and
maintaining the radar along the WFS.
\end{acknowledgments}

\end{article}


\begin{thebibliography}{28}
\providecommand{\natexlab}[1]{#1} \expandafter\ifx\csname
urlstyle\endcsname\relax
  \providecommand{\doi}[1]{doi:\discretionary{}{}{}#1}\else
  \providecommand{\doi}{doi:\discretionary{}{}{}\begingroup
  \urlstyle{rm}\Url}\fi

\bibitem[{\textit{Backer et~al.}(2003)}]{Backer-etal-03}
Backer, L.~C., et~al. (2003), Recreational exposure to aerosolized
brevetoxins
  during {F}lorida red tide events, \textit{Harmful Algae}, \textit{2}, 19--28.

\bibitem[{\textit{Bossart et~al.}(1998)\textit{Bossart, Baden, Ewing, Roberts,
  and Wright}}]{Bossart-etal-98}
Bossart, G.~D., D.~G. Baden, R.~Y. Ewing, B.~Roberts, and S.~D.
Wright (1998),
  Brevetoxicosis in manatees (\emph{Trichechus manatus} {L}atirostris) from the
  1996 epizoodic: gross, histopathologic and immunocytochemical features,
  \textit{Toxicol. Pathol.}, \textit{26}, 276--282.

\bibitem[{\textit{Brand and Guillard}(1981)}]{Brand-Guillard-81}
Brand, L.~E., and R.~R.~L. Guillard (1981), The effects of
continuous light and
  light intensity on the reproduction rates of twenty-two species of marine
  phytoplankto, \textit{J. Exp. Mar. Biol. Ecol.}, \textit{50}, 119--132.

\bibitem[{\textit{Chan}(1978)}]{Chan-78}
Chan, A.~T. (1978), Comparative physiological study of marine
diatoms and
  dinoflagellates in relation to irradiance and cell size. {I}. {G}rowth under
  continuous light, \textit{J. Phycol.}, \textit{14}, 396--402.

\bibitem[{\textit{Chassignet et~al.}(2006{\natexlab{a}})\textit{Chassignet,
  Hurlburt, Smedstad, Halliwell, Hogan, Wallcraft, Baraille, and
  Bleck}}]{Chassignet-etal-06b}
Chassignet, E.~P., H.~E. Hurlburt, O.~M. Smedstad, G.~R. Halliwell,
P.~J.
  Hogan, A.~J. Wallcraft, R.~Baraille, and R.~Bleck (2006{\natexlab{a}}), The
  {HYCOM (HYbrid Coordinate Ocean Model)} data assimilative system, \emph{J.
  Mar. Sys.}\textrm{, in press}.

\bibitem[{\textit{Chassignet et~al.}(2006{\natexlab{b}})\textit{Chassignet,
  Hurlburt, Smedstad, Halliwell, Hogan, Wallcraft, and
  Bleck}}]{Chassignet-etal-06a}
Chassignet, E.~P., H.~E. Hurlburt, O.~M. Smedstad, G.~R. Halliwell,
P.~J.
  Hogan, A.~J. Wallcraft, and R.~Bleck (2006{\natexlab{b}}), Ocean prediction
  with the {HYbrid Coordinate Ocean Model (HYCOM)}, in \textit{Ocean Weather
  Forecasting: An Integrated View of Oceanography}, edited by E.~P. Chassignet
  and J.~Verron, pp. 413--436, Springer.

\bibitem[{\textit{Dortch et~al.}(1998)\textit{Dortch, Moncreiff, Mendenhall,
  Parsons, Franks, and Hemphill}}]{Dortch-etal-98}
Dortch, Q., C.~Moncreiff, W.~Mendenhall, M.~Parsons, J.~Franks, and
K.~Hemphill
  (1998), Spread of \emph{{G}ymnodinium breve} into the northern {G}ulf of
  {M}exico, in \textit{Harmful Algae}, edited by B.~Reguera, J.~Blanco, M.~L.
  Fernandez, and T.~Wyatt, pp. 143--144, Xunta de Galicia and Intergovernmental
  Oceanographic Commission of UNESCO.

\bibitem[{\textit{Flewelling et~al.}(2005)}]{Flewelling-etal-05}
Flewelling, L.~J., et~al. (2005), Brevetoxicosis: Red tides and
marine mammal
  mortalities, \textit{Nature}, \textit{435}, 755--756.

\bibitem[{\textit{Geesey and Tester}(1993)}]{Geesey-Tester-93}
Geesey, M., and P.~A. Tester (1993), \emph{Gymnodinium breve}:
ubiquitous in
  {Gulf of Mexico} waters?, in \textit{Toxic Phytoplankton Blooms in the Sea},
  edited by S.~T. J. and Y.~Shimizu.

\bibitem[{\textit{Haller}(2000)}]{Haller-00}
Haller, G. (2000), Finding finite-time invariant manifolds in
two-dimensional
  velocity fields, \textit{Chaos}, \textit{10}, 99--108.

\bibitem[{\textit{Haller}(2001{\natexlab{a}})}]{Haller-01a}
Haller, G. (2001{\natexlab{a}}), Distinguished material surfaces and
coherent
  structures in 3{D} fluid flows, \textit{Physica D}, \textit{149}, 248--277.

\bibitem[{\textit{Haller}(2001{\natexlab{b}})}]{Haller-01b}
Haller, G. (2001{\natexlab{b}}), Lagrangian structures and the rate
of starin
  in a partition of two-dimensional turbulence, \textit{Phys. Fluids},
  \textit{13}, 3365--3385.

\bibitem[{\textit{Haller}(2002)}]{Haller-02}
Haller, G. (2002), Lagrangian coherent structures from approximate
velocity
  data, \textit{Physics Fluid}, \textit{14}, 1851--1861.

\bibitem[{\textit{Haller and Yuan}(2000)}]{Haller-Yuan-00}
Haller, G., and G.~Yuan (2000), Lagrangian coherent structures and
mixing in
  two-dimensional turbulence, \textit{Physica D}, \textit{147}, 352--370.

\bibitem[{\textit{Kirkpatrick et~al.}(2004)}]{Kirkpatrick-etal-04}
Kirkpatrick, B., et~al. (2004), Literature review of {F}lorida red
tide:
  Implications for human health effects, \textit{Harmful Algae}, \textit{3},
  99--115.

\bibitem[{\textit{Kusek et~al.}(1999)\textit{Kusek, Vargo, and
  Steidinger}}]{Kusek-etal-99}
Kusek, K.~M., G.~Vargo, and K.~Steidinger (1999),
\emph{{G}ymnodinimum breve}
  in the field, in the lab, and in the newspaper-a scientific and journalistic
  analysis of {F}lorida red tides, \textit{Contri. Mar. Sci.}, \textit{34},
  1--229.

\bibitem[{\textit{Landsberg}(2002)}]{Landsberg-02}
Landsberg, J.~H. (2002), The effects of harmful algal blooms on
aquatic
  organisms, \textit{Rev. Fish. Sci.}, \textit{10}, 113--390.

\bibitem[{\textit{Landsberg and Steidinger}(1998)}]{Landsberg-Steidinger-98}
Landsberg, J.~H., and K.~A. Steidinger (1998), A historical review
of
  \emph{{G}ymnodinium breve} red tides implicated in mass mortalities of the
  manatee (\emph{{T}richechus manatus} {L}atirostris) in {F}lorida, {USA}, in
  \textit{Harmful Algae}, edited by B.~Reguera, M.~L. Fernandez, and T.~Wyatt,
  pp. 97--100, Xunta de Galicia and Intergovernmental Oceanographic Commission
  of UNESCO.

\bibitem[{\textit{Lekien et~al.}(2005)\textit{Lekien, Coulliette, Mariano,
  Ryan, Shay, Haller, and Marsden}}]{Lekien-etal-05}
Lekien, F., C.~Coulliette, A.~J. Mariano, E.~H. Ryan, L.~K. Shay,
G.~Haller,
  and J.~E. Marsden (2005), Pollution release tied to invariant manifolds: {A}
  case study for the coast of {F}lorida, \textit{Physica D}, \textit{210},
  1--20.

\bibitem[{\textit{Magana et~al.}(2003)\textit{Magana, Contreras, and
  Villareal}}]{Magana-etal-03}
Magana, H.~A., C.~Contreras, and T.~A. Villareal (2003), A
historical
  assessment of \emph{Karenia brevis} in the western {Gulf of Mexico},
  \textit{Harmful Algae}, \textit{2}, 163--171.

\bibitem[{\textit{Morey et~al.}(2003)\textit{Morey, Martin, O'Brien, Wallcraft,
  and Zavala-Hidalgo}}]{Morey-etal-03}
Morey, S.~L., P.~J. Martin, J.~J. O'Brien, A.~A. Wallcraft, and
  J.~Zavala-Hidalgo (2003), Export pathways for river discharged fresh water in
  the northern {G}ulf of {M}exico, \textit{J. Geophys. Res.},
  \textit{108}(C10), 3303, \doi{10.1029/2002JC001674}.

\bibitem[{\textit{Shadden et~al.}(2005)\textit{Shadden, Lekien, and
  Marsden}}]{Shadden-etal-05}
Shadden, S.~C., F.~Lekien, and J.~E. Marsden (2005), Definition and
properties
  of {L}agrangian coherent structures from finite-time {L}yapunov exponents in
  two-dimensional aperidic flows, \textit{Physica D}, \textit{212}, 271--304.

\bibitem[{\textit{Shay et~al.}(2006)\textit{Shay, Martinez-Pedraja, Cook, Haus,
  and Weisberg}}]{Shay-etal-06}
Shay, L.~K., J.~Martinez-Pedraja, T.~M. Cook, B.~K. Haus, and R.~H.
Weisberg
  (2006), Surface current mapping using {W}ellen radars, \emph{J. Atmos.
  Oceanogr. Thech.} \textrm{, in press}.

\bibitem[{\textit{Shumway et~al.}(2003)\textit{Shumway, Allen, and
  Boersma}}]{Shumway-etal-03}
Shumway, S.~E., S.~M. Allen, and P.~D. Boersma (2003), Marine birds
and harmful
  algal blooms: sporadic victims or under-reported events?, \textit{Harmful
  Algae}, \textit{2}, 1--17.

\bibitem[{\textit{Steidinger and Haddad}(1981)}]{Steidinger-Haddad-81}
Steidinger, K.~A., and K.~Haddad (1981), Biologic and hydrographic
aspects of
  red-tides, \textit{BioScience}, \textit{31}, 814--819.

\bibitem[{\textit{Tester and Steidinger}(1997)}]{Tester-Steidinger-97}
Tester, P.~A., and K.~A. Steidinger (1997), \emph{Gymnodinium breve}
red tide
  blooms: Initiation, transport, and consequences of surface circulation,
  \textit{Limnol. Oceanogr.}, \textit{42}, 1039--1051.

\bibitem[{\textit{Wilson and Ray}(1956)}]{Wilson-Ray-56}
Wilson, W.~B., and S.~M. Ray (1956), The occurrence of
\emph{Gymnodinium breve}
  in the western {Gulf of Mexico}, \textit{Ecol.}, \textit{37}, 388.

\bibitem[{\textit{Yang et~al.}(1999)\textit{Yang, Weisberg, Niiler, Sturges,
  and Johnson}}]{Yang-etal-99}
Yang, H., R.~H. Weisberg, P.~P. Niiler, W.~Sturges, and W.~Johnson
(1999),
  Lagrangian circulation and forbidden zone on the {West Florida Shelf},
  \textit{Cont. Shelf. Res.}, \textit{19}, 1221--1245.

\end{thebibliography}
\end{document}